\documentstyle[aps,amssymb,amsfonts,times,mathptm,12pt]{revtex}
\begin{document}
\small
\normalsize
\renewcommand{\baselinestretch}{1}
\small
\normalsize
\title{Temporal Quantum Theory}
\author{Oliver Rudolph \thanks{email: o.rudolph@ic.ac.uk}}
\address{Theoretical Physics Group, Blackett Laboratory,
Imperial College of Science, Technology and
Medicine, \\ Prince Consort Road, London SW7 2BZ, United Kingdom}
\maketitle
\begin{abstract}
\noindent We propose a framework for temporal quantum theories
for the purpose of describing
states and observables associated with extended regions
of space time quantum mechanically. The proposal is motivated
by Isham's history theories. We discuss its relation
to Isham's history theories and
to standard quantum mechanics.
We generalize the Isham-Linden information entropy to the
present context.
\end{abstract}
\section{Introduction}
Standard nonrelativistic
quantum mechanics is based on the notion of state and
observable at fixed instants of time. In Hilbert space quantum mechanics
the states at some fixed time are represented by positive trace
class operators with trace one on some Hilbert space ${\frak H}$
and the observables
are identified with self-adjoint operators on ${\frak H}$.
The time evolution is governed by a semigroup $\{ U(t', t) \}$
of unitary operators on ${\frak H}$.

It has been felt by several authors that the notions of observables and
states at a fixed time slice are idealizations and might be
inappropriate when it comes to describing relativistic situations
quantum mechanically. For instance, in the algebraic approach to quantum
field theory, the theory is intrinsically
characterized by associating to every open
region ${\cal O}$ of space time an algebra ${\cal A}({\cal O})$
of operators on some Hilbert space
${\cal O} \to {\cal A}({\cal O})$ \cite{Haag96}.
Hegerfeldt's works \cite{Hegerfeldt89}
about localization observables are another example for
papers studying observables associated with bounded regions
of space time.
However, the notion of an observable
associated with an extended region in space time is foreign to
the conceptual framework and formalism of standard quantum mechanics.
Therefore, \emph{a priori} it is not clear whether and, if
so, how the formalism of quantum mechanics has to be changed to
include such space time observables associated with extended
regions.

A downright investigation to derive the possible structure of a
space time
quantum theory (and in particular the possible notions of temporal state
and observable) from the mathematical structure of standard quantum
mechanics seems to have first been undertaken by Isham
\cite{Isham94} who laid down a set of axioms for \emph{history
quantum theories}. With his history quantum theories Isham
pointed out an intrinsically quantum mechanical formalism dealing with
space time observables and states. The main paradigm of the approach is
to describe space time observables and states by operators on certain
tensor product Hilbert spaces. This is very natural and
works fine as long as we consider
finite dimensional Hilbert spaces and a discrete set of time points, but
when one takes into account infinite dimensional Hilbert spaces or
infinitely many time points or continuous time,
the tensor product paradigm is quite unnatural from a mathematical point of
view. In particular,
the decoherence functional, which represents the state in
Isham's approach, is in general a mathematically
unsatisfactorily behaving object (this will be discussed
in more detail in Section II below). Another problem of
this approach is that
the only presently known concrete example for a \emph{history quantum
theory} is
standard quantum mechanics over a finite dimensional Hilbert space.

In the present paper we take a fresh look upon the problem.
The target of the present investigation is to understand the
physical significance of the difficulties within the mathematical
framework in Isham's programme
more properly and to put forward a novel, mathematically more natural
framework for
space time quantum theories which
on the one hand is broad enough to embrace Isham's approach in the finite
dimensional case, but on the other hand goes significantly beyond it.

The paper is
organized as follows: in Section II we give an account of Isham's programme.
In Section III we shall put forward
our new mathematical framework for space time quantum mechanics and
show that our framework contains all known examples
of well defined \emph{general history quantum theories} as a subclass.
We shall also discuss in which sense the framework covers standard
quantum mechanics in the
infinite dimensional case. In Section IV we will discuss
the definition of an information entropy
for space time quantum theory which is a generalization of the Isham-Linden
information entropy to our approach.
Throughout this work we use the following notation: ${\frak H}$
always denotes the single time Hilbert space in ordinary quantum
mechanics, ${\frak K}$ always denotes the `propositions' Hilbert
space introduced in Section III.
${\cal H}$ or ${\cal V}$ denote general Hilbert
spaces or tensor product Hilbert spaces. The set of bounded
operators on some Hilbert space ${\cal H}$ is denoted by
${\cal B}({\cal H})$, the set of compact operators on ${\cal H}$
by ${\cal K}({\cal H})$ and the set of projection operators on
some Hilbert space ${\cal H}$ by
${\cal P}({\cal H})$. We adopt the convention that all inner
products and sesquilinear forms on Hilbert spaces are linear in
their second variable and conjugate linear in the first variable.

\section{History quantum theories}
\subsubsection*{Generalities}
In the mathematical formulation of standard quantum mechanics every
quantum mechanical system is characterized by some Hilbert space
${\frak H}$ and some
semigroup of unitary time evolution operators
acting on ${\frak H}$. The possible states of
the quantum mechanical system at some fixed instant of time are
identified with the trace class operators on ${\frak H}$ and the
possible observables are identified with the self-adjoint operators on
${\frak H}$. It is well known that
according to the spectral theorem every observable can be disintegrated
into so called yes-no observables represented by projection operators on
${\frak H}$. The projection operators
represent the elementary propositions about the system.
The quantum mechanical probability for the proposition represented by
the projection $P$ is in the state $\rho$ then given by
$\text{tr}_{{\frak H}}(P \rho)$.

As already anticipated above, in the history approach one
aims at including space time observables in the quantum mechanical
formalism. In a first moderate step one considers a finite sequence
of projection operators $P_{t_1},  \cdots, P_{t_n}$ associated with
times $t_1, \cdots, t_n$ which corresponds to a time sequence of
propositions. Such a sequence is called a homogeneous history.
The quantum mechanical probability of a history $h \simeq \{ P_{t_1},
\cdots, P_{t_n} \}$ is given by $\text{tr}_{{\frak H}}(P_{t_n} \cdots
P_{t_1} \rho P_{t_1} \cdots P_{t_n})$ [notice that we work in the
Heisenberg picture and suppress for notational simplicity the time
dependence of the operators].
Slightly abstracting from this expression one defines the
\emph{decoherence functional} $d_\rho$ on pairs of homogeneous histories by
\cite{Griffiths84,Omnes94}
\begin{equation} \label{decf1}
d_\rho(h,k) := \text{tr}_{{\frak H}}(P_{t_n} \cdots P_{t_1} \rho
Q_{t_1} \cdots Q_{t_n}), \end{equation} where $h \simeq \{ P_{t_1},
\cdots, P_{t_n} \}$ and $k \simeq \{ Q_{t_1},
\cdots, Q_{t_n} \}$. Histories which differ from each other only by the
insertion or the omission of the unit operator at intermediate times are
physically equivalent and identified which each other.

The next major step in the construction of a \emph{general history
theory} is to embed
the set of all (equivalence classes of)
homogeneous histories injectively into a larger space
${\cal V},$ which is endowed with a partially defined sum,
such that the decoherence functional $d_\rho$ can be
extended unambiguously to a bounded bi-additive
functional $D_\rho : {\cal V} \times
{\cal V} \to {\Bbb C}$ subject to the further conditions
(i) $D_\rho(u,v) = D_\rho(v,u)^*$ for all $u,v \in
{\cal V}$ and (ii) $D_\rho(u,u) \geq 0$ for all $u \in {\cal V}$.
[In the previous literature about \emph{general history quantum
theories} the space ${\cal V}$
carried usually
some additional structure, e.g., that of a lattice or a
unital *-algebra. In the history
approach put forward by Gell-Mann and Hartle (see
\cite{GellMann90a} - \cite{GellMann93} and below)
the embedding of the
homogeneous histories into the larger space of so called \emph{class
operators} is not injective.]
The
homogeneous histories are identified with their images in ${\cal V}$
which are also called homogeneous histories.
The elements in ${\cal V}$
which are not an image of some homogeneous history are called
\emph{inhomogeneous histories}. All elements in ${\cal V}$ are
interpreted as the general (measurable) space time propositions in
the theory.

A subset ${\cal V}_0$ of ${\cal V}$ is called a \emph{consistent set
of histories} if $D_\rho$ induces a probability measure $p : {\cal V}_0
\to {\Bbb C}, p(v_0) :=
D_\rho(v_0, v_0)$ on ${\cal V}_0$. The consistent subsets of
${\cal V}$ are exactly those subsets ${\cal V}_0$ which can be
endowed with a Boolean structure and which satisfy $\text{Re} \:
\, D_\rho(u_0, v_0) =0$ for all mutually disjoint
$u_0, v_0 \in {\cal V}_0$. The quantum
character of the theory and the principle of complementarity exhibit
themselves in the fact that there are several mutually inconsistent
consistent subsets of ${\cal V}$ \cite{Omnes94}.

\subsubsection*{Isham's history theories}
In the history approach developed by Isham and coworkers
\cite{Isham94}-\cite{IshamLSS97},
\cite{Schreckenberg95}-\cite{Schreckenberg96a} and other authors
\cite{Pulmannova95}-\cite{RudolphW98},
\cite{Wright95}-\cite{Wright97b}
the crucial
observation was that homogeneous histories as above can be
mathematically conveniently
described by using a tensor product formalism.
Correspondingly, the homogeneous
history $h \simeq \{P_{t_1}, \cdots, P_{t_m} \}$ is mapped to the
projection operator $P_{t_1} \otimes \cdots \otimes P_{t_m}$
on the tensor product Hilbert space $\otimes_{ t_i \in
\{t_1, \cdots, t_m \} }
{\frak H}_{t_i}$ where ${\frak H}_{t_i} = {\frak H}$ for
all $i.$
It is mathematically convenient to postulate that the space of all
histories is given by ${\cal P}(\otimes_{ t_i \in \{ t_1, \cdots, t_n \}}
{\frak H}_{t_i})$.
Some of the \emph{inhomogeneous} elements
within ${\cal P}(\otimes_{ t_i \in \{ t_1, \cdots, t_n \}}
{\frak H}_{t_i})$ can be straightforwardly interpreted as
coarse grainings of homogeneous histories (in the proposition
picture these histories represent composed propositions such as ``$h_1$
or $h_2$ are true'' etc.)
However, if the single time Hilbert space ${\frak H}$ is
infinite dimensional there are always some elements in
${\cal P}(\otimes_{t_i \in \{ t_1, \cdots, t_n \}} {\frak H}_{t_i})$ which
admit no physical interpretation as coarse graining of homogeneous
histories.

Again histories which
differ from each other only by the
insertion or the omission of the unit operator at intermediate times are
physically equivalent and identified which each other. We shall refer to
this natural equivalence relation between histories as the
\emph{canonical equivalence} and use the symbol $\sim_c$ in the sequel.
For every history $h$ defined on $\otimes_{t_i \in \{t_1, \cdots, t_n
\}} {\frak H}_{t_i}$, where ${\frak H}_{t_i} = {\frak H}$ for all
$i$, consider its equivalence class $\varepsilon(h)$ of histories.
We say that a finite set of time points $s = \{ t_1, \cdots, t_m \}$
is the \emph{support of} $h$ if (i) there is
an element in $\varepsilon(h)$ defined on $\otimes_{t_i \in s}
{\frak H}_{t_i}$ and (ii) for every proper subset $s'$ of $s$ there
is no element in $\varepsilon(h)$ defined on $\otimes_{t_i \in s'}
{\frak H}_{t_i}$. Mathematically
speaking the system $\{ {\cal P}({\frak H}_{t_1} \otimes \cdots
\otimes {\frak H}_{t_n}) \vert \{t_1, \cdots, t_n \} \subset
{\Bbb R} \}$
of sets of projection operators associated with all possible finite sets
of time points forms a \emph{directed} or \emph{inductive system}.
The space of
all histories is identified with the disjoint union over the sets of
\emph{all} projections on all finite tensor product Hilbert spaces of the
form
$ \otimes_{ t_i \in \{t_1, \cdots, t_n \} } {\frak H}_{t_i}$
modulo the physical
equivalence $\sim_c$ or -- mathematically more
precise -- with the \emph{direct} or \emph{inductive limit} of the
directed system of histories $\{ {\cal P}({\frak H}_{t_1}
\otimes \cdots
\otimes {\frak H}_{t_n}) \vert \{t_1, \cdots, t_n \} \subset
{\Bbb R} \}$ (a proof for the existence of this direct limit
can for instance be found in \cite{Pulmannova95}).

In an important paper Isham, Linden and Schreckenberg \cite{IshamLS94}
showed that if the
single time Hilbert space ${\frak H}$ is finite dimensional, the
decoherence functional $d_\rho$ defined on pairs of homogeneous
histories can be
unambiguously extended to a bounded bi-additive
functional $D_\rho$ on the space of all histories.
Moreover, they showed that for fixed $n$ and $\rho$ there exists
a trace class
operator ${\frak X}_\rho$ on $\otimes_{2n} {\frak H}$ such that
$D_\rho$ can be written as \[ D_\rho(u,v) = \text{tr}_{\otimes_{2n}
{\frak H}} (u \otimes v {\frak X}_\rho), \]
for all $u,v \in {\cal P}({\frak H}_{t_1} \otimes \cdots
\otimes {\frak H}_{t_n}).$

Abstracting from this result the properties of general decoherence
functionals Isham \cite{Isham94}
arrived at an axiomatic characterization of general
history quantum theories according to which a general history quantum
theory is given by its space ${\cal U}$ of histories
and by its space ${\cal D}$ of
decoherence functionals. The histories in ${\cal U}$ and the
decoherence functionals in ${\cal D}$ represent the
(measurable) propositions and the states in the theory respectively.
The space of
histories ${\cal U}$
is required to have a partial sum defined on it and to contain
a unit $\mathbf{1}$. Every decoherence functional $d \in {\cal D}$
is required to
be bounded and additive in both arguments and has to satisfy
(i) $d(\mathbf{1}, \mathbf{1}) = 1$, (ii) $d(x,x) \geq 0$, (iii) $d(x,y) =
d(y,x)^*$, for all histories $x,y \in {\cal U}$.

A choice for the space of histories in a general history quantum theory
suggesting itself is the set of projection operators
${\cal P}({\cal H})$ on some Hilbert
space ${\cal H}$ or -- slightly more general -- the set of projection
operators ${\cal P}({\cal A})$ in a von Neumann algebra
${\cal A}$. In the case that the space of histories is given by
${\cal P}({\cal H})$ for some finite dimensional Hilbert space
${\cal H}$ Isham, Linden and Schreckenberg \cite{IshamLS94}
showed that for every bounded
decoherence functional $d$ on ${\cal P}({\cal H})$ there exists a
trace class operator ${\frak X}_d$
on ${\cal H} \otimes {\cal H}$ such that \begin{equation}
d(x,y) = \text{tr}_{{\cal H} \otimes {\cal H}}(x \otimes y
{\frak X}_d) \label{ILS2} \end{equation}
holds for all $x,y \in {\cal P}({\cal H})$.
Subsequently, Isham and coworkers studied different aspects of general
history quantum theories over finite dimensional Hilbert spaces in some
detail, buoying up the fruitfulness of the tensor product based
approach.

On the other hand, the use of tensor product spaces to describe
temporarily extended objects has its limitations. Despite some
interesting research and progress made recently
\cite{IshamL95,IshamLSS97}, the incorporation of
continuous histories into the approach
is still a challenge and the tensor product based
approach does not seem to be well adapted to it.

Recently Wright and the author \cite{RudolphW98} showed
that if the single time Hilbert
space ${\frak H}$ in standard quantum mechanics
is infinite dimensional, then no decoherence
functional $d_\rho$ (corresponding to the initial state $\rho$, see
Equation (\ref{decf1}))
defined on homogeneous histories can be extended
to a bounded, or even to a finitely valued functional on the space
of ``all histories'' in Isham's approach. In \cite{RudolphW98} an
example for an element $h_\infty \in
{\cal P}({\frak H}_{t_1} \otimes \cdots
\otimes {\frak H}_{t_n})$ was constructed such that no
decoherence functional $d_\rho$ assumes a finite value at
$h_\infty$, if extended (see also Appendix B).
One possible way out of this dilemma is to allow for
decoherence functionals assuming values in
the Riemann sphere ${\Bbb C} \cup \{ \infty \}$.
Histories $h$ with $d(h,h) = \infty$ are then called
\emph{singular} histories for the decoherence functional $d.$
Histories $h \in {\cal P}({\frak H}_{t_1} \otimes \cdots
\otimes {\frak H}_{t_n})$ with $d_\rho(h,h) > 1$
or $d_\rho(h,h) = \infty$ are in no consistent set and represent
no physical propositions in the state $d_\rho$. Adopting this point of view
one could simply forget about the singular histories. However, in
\cite{RudolphW98} it has been shown that there are certain
histories in ${\cal P}({\frak H}_{t_1} \otimes \cdots
\otimes {\frak H}_{t_n})$ which are singular for
every decoherence functional $d_\rho$. Thus these singular histories
will be in no consistent set of histories for all states $d_\rho.$
This result indicates that in the infinite dimensional
case the space ${\cal P}({\frak H}_{t_1} \otimes \cdots
\otimes {\frak H}_{t_n})$ contains unphysical
elements which represent no physical histories at all.
We shall propose an alternative mathematical framework for the
description of temporal quantum theories and we shall show that
the decoherence functional $d_\rho$ of the history version of standard
quantum mechanics (when cast into the new framework) is a
finitely valued functional.

For the convenience of the reader and for later reference we cite the
following representation of the standard decoherence functional as an in
general infinite sum.
For $n$-time homogeneous histories $p$ and $q$ and for a finite
dimensional or infinite dimensional single time Hilbert space
${\frak H}$ \cite{IshamLS94,RudolphW98} we have:
{\small
\begin{equation} \label{decf}
d_\rho(p,q) = \sum_{j_1, \cdots, j_{2n}}
\omega_{j_1}
\left\langle e^{2n}_{j_{2n}} \otimes \cdots \otimes e^{n+1}_{j_{n+1}}
\otimes \psi_{j_1} \otimes e^2_{j_2} \otimes \cdots \otimes e^n_{j_n},
(p \otimes q) (\psi_{j_1} \otimes e^{2n}_{j_{2n}} \otimes \cdots \otimes
e^{n+2}_{j_{n+2}} \otimes e^2_{j_2} \otimes \cdots \otimes
e^{n+1}_{j_{n+1}}) \right\rangle, \end{equation}}

\noindent where
$\{e^k_{j_k} \}$ are orthonormal bases of ${\frak H}$ for all $2
\leq k \leq 2n$, where $\rho = \sum_{i}
\omega_i P_{\psi_i}$ denotes the spectral resolution of
$\rho$ and $P_{\psi_i}$ denotes the projection operator onto the
subspace of ${\frak H}$ spanned by $\psi_i$, and where $\omega_i \geq
0$ for all $i$. We shall
always assume that the orthonormal system $\{ \psi_i \}$
has been extended to an orthonormal basis of $\frak H$.

\subsubsection*{Gell-Mann--Hartle history theories}
The main predecessor to Isham's history quantum theories was the
approach put forward by Gell-Mann and Hartle \cite{GellMann90a} -
\cite{GellMann93}. In this approach the
homogeneous histories are mapped to so called class operators \[ h
\simeq \{ P_{t_1}, \cdots, P_{t_n} \} \rightarrow C(h) := P_{t_1} \cdots
P_{t_n}. \] Notice that again
we use the Heisenberg picture and suppress the
explicit time dependence of the operators.
Class operators act on the single time Hilbert space ${\frak H}$.
The inhomogeneous histories are indirectly defined by so called
\emph{coarse graining} prescriptions. An inhomogeneous history in the
Gell-Mann--Hartle approach is in general a sum of class operators
which correspond to mutually exclusive homogeneous histories.
While the decoherence functional
$d_\rho$ extends to inhomogeneous class operators straightforwardly (by
linearity in both arguments) and the
interpretation of the inhomogeneous histories is equally
straightforward, the formalism essentially stays on the level of
homogeneous histories and we are lacking a simple and direct
characterization of the mathematical
structure of the space of class operators which makes the
Gell-Mann--Hartle approach
virtually incomprehensible to a rigorous mathematical investigation:
given an arbitrary
operator ${\frak c}$ in the unit sphere of ${\cal B}({\frak H})$
we have no
simple criterion to decide whether ${\frak c}$ is a class operator or
not and if so, whether it is a homogeneous class operator or an
inhomogeneous class operator. Moreover, as already mentioned above,
the map of homogeneous
histories to class operators is by no means injective and in general
there is more than one homogeneous history corresponding to a given
class operator.

\section{Temporal quantum mechanics}
\subsection{The general framework}
In this section we put forward our framework for space time quantum
mechanics. We first state the main principles without further motivation
and then proceed to show
how standard quantum mechanics and Isham's general
history theories over finite dimensional Hilbert spaces
fit into the scheme which serves as an \emph{a posteriori}
motivation.

The basic ingredient in our framework of \emph{temporal quantum theories}
is a Hilbert space ${\frak K}$ whose elements are interpreted as the
(measurable) local space time propositions about the system.
Actually not all elements in ${\frak K}$ represent
physically meaningful propositions, see below.
In spite of this we shall refer to
the Hilbert space
${\frak K}$ as the space of propositions and to
the elements of ${\frak K}$ simply as
propositions.
We assume that
there exists one distinguished element
in ${\frak K}$, denoted by $e$,
which represents the indifferent
proposition which is always true. The
trivial proposition complementary to $e$ which is always false is
identified with the zero vector in ${\frak K}$.

We shall argue that the norm induced by
the inner product $\langle \cdot, \cdot \rangle$ in ${\frak K}$
subsumes the \emph{a priori} structural
information about the propositions which is encoded
within the propositions Hilbert space ${\frak K}$. It is a
quantitative measure for the fine-grainedness of propositions within the
descriptive scheme provided by
${\frak K}$, i.e., the
smaller $\langle b, b \rangle$ is, the more ``fine grained'' is the
proposition corresponding to $b \in {\frak K}$.
More specifically,
we shall see below
that the amount of information associated with a proposition
$b \in {\frak K}$ is given by $- \ln \frac{p(b)}{\langle b, b \rangle}$
(where $p(b)$ denotes the probability of $b$) which is just the
difference between the information associated with the probability
distribution and the structural information encoded in $\langle \cdot,
\cdot \rangle$.

It is very important not to confuse the ``temporal''
Hilbert space ${\frak K}$ with the single time Hilbert space
${\frak H}$ in ordinary quantum mechanics. The elements of the single
time Hilbert space ${\frak H}$
in ordinary quantum mechanics correspond to the
possible pure \emph{states} of the system.
Hence, the two Hilbert spaces
${\frak K}$ and ${\frak H}$
 have \emph{a priori} nothing
to do with each other.
We shall clarify the \emph{a posteriori}
relation between the ``temporal''
Hilbert space ${\frak K}$ and the single time Hilbert space
${\frak H}$ below.

We shall see that in Isham's abstract history quantum theories
the propositions Hilbert space ${\frak K}$ in general
depends both on the single time Hilbert space ${\frak H}$ and on the
quantum state given by some decoherence functional $d$.
Physically this reflects the fact that the
global propositions one may sensible ask about the system can change when
the global state $d$ of the system is changed. We shall
see that essentially the Hilbert space ${\frak K}$ corresponding to
a decoherence functional $d$ is constructed
from a larger space of propositions
by omitting some propositions with vanishing probability in the state
$d$.

The quantum mechanical ``temporal'' states of the system are given by
self-adjoint bounded operators $T$ on
${\frak K}$ such that
$\langle e, T e \rangle = 1$, where $\langle \cdot, \cdot
\rangle$ denotes the inner product in ${\frak K}$.
We denote the set of
all such operators by ${\cal W}_e({\frak K})$.
Wright first showed that one can associate a bounded operator $T$
with every bounded decoherence functional in a
general history quantum theory (see
\cite{Wright95} and below). Thus
we shall refer to the state operator $T$
also as the \emph{Wright operator} of the system.

Since the bounded, self-adjoint operators on ${\frak K}$
are in one-to-one
correspondence with bounded sesquilinear forms on ${\frak K}$, we can
alternatively define
the states as bounded sesquilinear forms $s$
on ${\frak K}$ satisfying $s(e,e) =1$.

The expression for the probability density
for a certain proposition $x \in {\frak K}$ in the state
given by the Wright
operator $T \in {\cal W}_e({\frak K})$ is given by
\begin{equation}
\label{propa}
p_T(x) = \langle x, T x \rangle \end{equation}
for all $x \in {\frak K}$. To wit, the Wright operator $T$ should
be thought of as generating the probability
densities for local propositions.
Once the space time region is fixed, probabilities for local propositions
can be obtained from
Equation (\ref{propa}) by multiplying with
a suitable normalization factor
(representing the number of
quantum degrees of freedom within the region in question).
For example, we shall see that in
the history formulation of standard quantum mechanics,
discussed in Section III.B, we can find for every fixed temporal support
$s$ a $T$ with $\langle e, T e
\rangle = 1$ such that $p_T$ can be
interpreted as probability for propositions with temporal support $s$.
As will become clear below, in
general, however, the probabilities associated with propositions
corresponding to different temporal supports cannot be properly
normalized. Only the probability density and the probabilities for
fixed temporal supports can be
brought into the form (\ref{propa}) for some appropriate $T$.
This substantiates our proposal
that in a general theory without an {\em a priori}
space time the Wright operator generates the space time
probability densities via Equation (\ref{propa}).

>From the definition of the ``probability''
functional $p_T : {\frak K} \to
{\Bbb R}$ it is obvious that $p_T$ is not necessarily positive definite
and defines no linear functional on all
of ${\frak K}$. Thus it is useful to adopt a consistent-histories-type
point of view \cite{Omnes94}.
All propositions $y \in {\frak K}$ with either
$p_T(y) < 0$ or
$p_T(y) > 1$ are assumed to be physically meaningless in the state $T \in
{\cal W}_e({\frak K})$. We say that a set ${\cal C} :=
\{ x_i \vert i \in I, x_i \in {\frak K} \}$
is \emph{consistent} in the state $T \in
{\cal W}_e({\frak K})$ if (i)
$x_i \perp x_j$ for $i \neq
j$, (ii) $\sum_i x_i = e$, (iii)
$0 < p_T(x_i) \leq 1$ for all $i$, and (iv)
$\sum_i p_T(x_i) = p_T \left( \sum_i x_i \right) = p_T(e) = 1$.
Consistent sets of propositions represent the analogue of sets of
commuting, mutually exclusive
yes-no observables in standard quantum mechanics and the existence of
several mutually inconsistent (complementary)
consistent sets of propositions unravels
the quantum character of the theory.
The space of \emph{all}
(measurable) propositions in our approach carries the
structure of a Hilbert space but no lattice theoretical structure. Only
the consistent subsets of ${\frak K}$ carry the structure of a
Boolean algebra. It is a virtue of the consistent histories philosophy
that it allows us to consider spaces of propositions without any lattice
theoretical structure on it and our unifying mathematical treatment of
temporal quantum theories relies heavily upon this feature of the
consistent histories philosophy.
Consistent sets of propositions are
also called \emph{windows} or
\emph{frameworks} for the description
of a quantum system. A \emph{refinement} $W_2 := \{ y_j \}_{j \in J}$
of a consistent set  $W_1 := \{ x_i \}_{i \in I}$ for $T$ is a
consistent set for $T$ such that each element $x_i \in W_1$ can be
written as a finite sum of elements in $W_2$. A consistent set is said
to be \emph{maximally refined}
if it has no consistent refinement. \\

We call ${\frak K}$ the space of propositions about the system.
However, notice that there are in general many elements in
${\frak K}$ which are in no consistent set for some state $T$.
The Hilbert space ${\frak K}$
serves as a
mathematically nice space into which the propositions are
embedded. The example of standard quantum mechanics
discussed in the next subsection will clarify this point.
The consistency condition for every physical state
singles out the elements
which can be interpreted as physically meaningful propositions in
the respective state of the system.

The framework for temporal quantum theories introduced above must not be
considered as a fixed, rigid set of axioms but should be
viewed as a set of \emph{cum grano salis}
working hypotheses which might be in need for change in the future.
In fact we shall see that the history version of standard quantum
mechanics over infinite dimensional Hilbert spaces
fits only into the framework when one allows for elements of
infinite norm in the Hilbert space ${\frak K}$. We shall call such
Hilbert spaces \emph{improper Hilbert spaces}, see Appendix A.
[We mention that there
is an alternative formulation of the history version of
standard quantum mechanics over an infinite dimensional Hilbert
space as a temporal quantum theory obtained in \cite{RudolphW98}
in which all information about
probabilities and the quantum state is thrown into the
propositions Hilbert space $\frak K$.]
At the basis of the present investigation is the (tacit)
assumption that essential features of a mathematical
framework for temporal quantum
mechanics can be read off a temporal reformulation of ordinary quantum
mechanics.
Infinite dimensional Hilbert spaces
are needed in ordinary quantum mechanics
to describe observables with continuous spectrum like
position and momentum observables. Thus it can be argued that
the mathematical difficulties
in the history version of quantum mechanics over infinite
dimensional Hilbert spaces are connected with the fact that
standard quantum mechanics
involves the concept of an underlying space time continuum (or
involves other observables with continuous spectrum).
An appealing idea put forward by many authors is that
in a quantum theory of space time observables the underlying
concept of space time should -- in one way or another --
be of a discrete nature.
Moreover, the form of the
canonical decoherence functional in standard
quantum mechanics is
based on the idealized notion of state and observable at a
fixed time instant and is
intrinsically non-relativistic (in that the
prescription for the computation of probabilities involves a series of
\emph{pro forma} global reductions of the state). Accordingly
we cannot expect that the
mathematical structure of a quantum theory of space time events can be
fully derived from standard quantum mechanics.
In particular we feel that the appearance of `infinitely' coarse grained
histories in the history version of standard quantum mechanics is a
reflection of the fact
that the theory is based on overidealized notions like observables
and states at some instant of time. Therefore, arguably, it is
inappropriate to base our
mathematical framework for temporal quantum theories upon the concept of
improper Hilbert space.

\subsection{Examples}
\subsubsection*{Isham's history quantum theories}
As a first example we consider Isham's abstract
history quantum theories over
some finite or infinite dimensional Hilbert space ${\cal V}$ (with
dimension greater than 2). In this
approach the space of histories is identified with the set
${\cal P}({\cal V})$ of projections
on ${\cal V}$ and the state is given by some bounded
decoherence functional $d$, i.e., by some bounded bi-orthoadditive
functional
$d : {\cal P}({\cal V}) \times {\cal P}({\cal V}) \to
{\Bbb C}$ satisfying (i) $d(1,1) = 1$, (ii) $d(p,q) = d(q,p)^*$ and
(iii) $d(p,p) \geq 0$, for all $p,q \in {\cal P}({\cal V})$.

We appeal now to an important result of Wright \cite{Wright95}, Corollary
4, according to which there exists a Hilbert space ${\frak K}$, a
self-adjoint bounded operator $T$ on ${\frak K}$ and a map $x \to [x]$
from ${\cal B}({\cal V})$ into a dense subspace of ${\frak K}$
such that $D : {\cal B}({\cal V}) \times {\cal B}({\cal V})
\to {\Bbb C}, D(x,y) = \langle [x], T [y] \rangle$ is an extension of $d$.
With $e := [1]$ it follows $\langle e, T e \rangle =1$. Thus Wright's
result implies that a general history quantum theory can always be
brought into the form of a space time quantum theory.

If ${\cal V}$ is finite dimensional, then it is possible to show that
the Hilbert space ${\frak K}$ can be chosen independently of $d$ and
may be identified with ${\cal B}({\cal V})$, see Remark (v) in
Section III in \cite{Wright95}.

Wright's result depends crucially on the fact that $D:
{\cal B}({\cal V}) \times {\cal B}({\cal V}) \to {\Bbb C}$
satisfies a Haagerup-Pisier-Grothendieck inequality, i.e., that
there exists a positive linear functional $\phi$
on ${\cal B}({\cal V})$ with $\phi(1) = 1$ and a constant $C>0$
such that \[ \vert D(x,y) \vert^2 \leq C \phi(x x^\dagger +
x^\dagger x) \phi(y y^\dagger + y^\dagger y), \] for all $x,y \in
{\cal B}({\cal V})$. The semi inner product on
${\cal B}({\cal V})$ is then constructed from $\phi$ as $\langle y,
x \rangle_\phi = \frac{1}{2} \phi(x y^\dagger + y^\dagger x)$.
Let $N_\phi$ be the corresponding null space,
then $\frak K$ is chosen as the completion of ${\cal B}({\cal V})/N_\phi$
with respect to the inner product induced by $\phi$.
The functional $\phi$ is not unique. For every $\phi$ there exists
a trace class operator $\tau_\phi$ on ${\cal V}$ such that $\phi(x) =
\text{tr}_{{\cal V}}(x \tau_\phi)$ for all $x \in
{\cal B}({\cal V})$.
Thus the Hilbert space ${\frak K}$ depends on $d$. As just explained
the space ${\frak K}$ is always (the completion
of a space) of the form ${\cal B}({\cal V})/N$ where
$N$ is a set of elements of ${\cal B}({\cal V})$ with $D(n,n) = 0$
for $n \in N$. Loosely speaking one may think of ${\frak K}$ as a
subspace of ${\cal B}({\cal V})$ in which some unphysical elements
with vanishing ``probability'' have been dismissed.
The different
choices of $\phi$ correspond to different null spaces $N_\phi$.
The probabilities
for physical propositions do not change for different choices
of $\phi$ but the number of unphysical propositions with
probability zero in ${\frak K}$ changes.  That is
changing $\phi$ amounts to changing the number of redundant
propositions in ${\frak K}$.

Wright's result also applies to arbitrary von Neumann algebras with no
type $I_2$ direct summand.

\subsubsection*{The history version of standard quantum mechanics}
As before we denote the single time Hilbert space by ${\frak H}$ and
the decoherence functional associated with the state $\rho$ by $d_\rho$.

Consider first the case that ${\frak H}$ is finite dimensional.
As discussed above the homogeneous histories associated with the times
$\{t_1, \cdots, t_n \}$ are identified with homogeneous projection
operators of the form $P_{t_1} \otimes \cdots \otimes P_{t_n}$ on
${\frak H}_{t_1} \otimes \cdots \otimes {\frak H}_{t_n}$.
The space of
all histories is identified with the direct limit of the
directed system of histories $\{ {\cal P}({\frak H}_{t_1}
\otimes \cdots
\otimes {\frak H}_{t_n}) \vert \{t_1, \cdots, t_n \} \subset
{\Bbb R} \}$ as discussed earlier.
Consider some fixed $n$ and some fixed set of times $\{ t_1, \cdots, t_n
\}$ and write ${\cal V}_{t_1, \cdots, t_n} :=
{\frak H}_{t_1}
\otimes \cdots \otimes {\frak H}_{t_n}$. Consider
the restriction of $d_\rho$ to ${\cal P}({\cal V}_{t_1, \cdots, t_n})$.
Isham, Linden and
Schreckenberg have shown that there exists a trace class operator
${\frak X}_d$ on ${\cal V}_{t_1, \cdots, t_n} \otimes
{\cal V}_{t_1, \cdots, t_n}$ such that \[ d_\rho(p,q) =
\text{tr}_{{\cal V}_{t_1, \cdots, t_n} \otimes
{\cal V}_{t_1, \cdots, t_n}}(p
\otimes q {\frak X}_d), \] for all $p,q \in
{\cal P}({\cal V}_{t_1, \cdots, t_n})$.
>From this it is obvious that $d_\rho$ can be extended to a bounded
functional on all of
${\cal B}({\cal V}_{t_1, \cdots, t_n})$. For what follows
it is convenient to
introduce the \emph{density} \[ \delta_\rho(p,q) :=
\frac{\text{tr}_{{\cal V}_{t_1, \cdots, t_n} \otimes
{\cal V}_{t_1, \cdots, t_n}}(p
\otimes q {\frak X}_d)}{\text{tr}_{{\cal V}_{t_1, \cdots, t_n}}(1)}, \]
for all $p,q \in
{\cal P}({\cal V}_{t_1, \cdots, t_n})$. The quantity
$\delta_\rho(p,p)$ is then a probability per quantum (space
time) degree of freedom.

One can look upon $d_\rho$ and the space of all histories from a
slightly different perspective. To this end consider the directed system
$\{ {\cal B}({\frak H}_{t_1}
\otimes \cdots
\otimes {\frak H}_{t_n}) \vert \{t_1, \cdots, t_n \} \subset
{\Bbb R} \}$ and its direct limit which we denote by ${\cal B}$
(the existence of this direct limit as a $C^*$-algebra follows,
e.g., from Proposition 11.4.1 in \cite{KadisonR86}).
Consider the map $\pi$ which maps every homogeneous bounded operator
$b:= b_{t_1} \otimes \cdots \otimes b_{t_n}$ on
${\cal B}({\cal V}_{t_1, \cdots, t_n})$
to $\pi(b) := b_{t_1}
\cdots b_{t_n} \in {\cal B}({\frak H})$. From Proposition 11.1.8
(ii) in \cite{KadisonR86} it follows
that $\pi$ can be uniquely extended to a
linear map from ${\cal B}$ to ${\cal B}({\frak H})$, which we
will also denote by $\pi$.
Every $d_\rho$ can be extended  to a sesquilinear form $D_\rho$
on ${\cal B}$
such that this
extension can then be written
as \[ D_\rho : {\cal B} \times {\cal B} \to {\Bbb C},
D_\rho(b_1, b_2) := \text{tr}_{{\frak H}}(\pi(b_1)^\dagger
\rho \pi(b_2)). \]
The corresponding extension of $\delta_\rho$ will be denoted by
$\Delta_\rho.$
Consider some fixed set of time points $\{t_1, \cdots, t_n \}$. We
define an inner product on ${\cal B}$ by \[ \langle b_1, b_2
\rangle
:= \frac{\text{tr}_{{\frak H}_{t_1} \otimes \cdots \otimes
{\frak H}_{t_n}}(b_1^\dagger b_2)}{\text{tr}_{{\frak H}_{t_1}
\otimes \cdots \otimes
{\frak H}_{t_n}}(1)}, \] for
all $b_1, b_2 \in {\cal B}$ where $\{ t_1, \cdots, t_n \}$ denotes the
support of $b_1^\dagger b_2$. We shall denote the norm induced by
the inner product
$\langle \cdot, \cdot \rangle$ by $\Vert \cdot \Vert_2$ for
reasons to become clear below.
The factor $\text{tr}_{{\frak H}_{t_1}
\otimes \cdots \otimes
{\frak H}_{t_n}}(1)$ in needed to ensure the additivity of
$\langle \cdot, \cdot \rangle$ on all of ${\cal B}$. Denote by
${\frak K}$ the Hilbert space completion of ${\cal B}$ with
respect to $\langle \cdot, \cdot \rangle$. Notice that although
${\frak H}$ is finite dimensional, ${\cal B}$ and ${\frak K}$
are infinite dimensional. Since the trace is a bounded linear functional on
${\cal B}({\frak H}_{t_1} \otimes \cdots \otimes
{\frak H}_{t_n})$ and since $D_\rho$ is bounded with respect to the
ordinary operator norm, it follows that
${\Delta}_\rho$ extends uniquely to a bounded sesquilinear form on
${\frak K}$. So there exists a bounded,
self-adjoint operator $\tau_\rho$ in
${\cal B}({\frak K})$ such that $\Delta_\rho(x,y) = \langle
x, \tau_\rho y \rangle$ for all $x, y \in {\frak K}$. Let $b \in {\cal B}$
and let $\{ t_1, \cdots, t_m \}$ denote the temporal support of $b$.
Define \[ T_\rho b := (\dim {\frak H})^{m} \tau_\rho b. \] Then $T_\rho$
is an unbounded operator on $\frak K$ whose (dense) domain of definition
is $\cal B$. $T_\rho$ is not self-adjoint on all of $\cal B$. However,

the restriction of $T_\rho$
to a subset of $\cal B$ containing only elements with fixed support
is bounded and self-adjoint.
Then the sesquilinear form $D_\rho$ obviously satisfies
$D_\rho(b, b) = \langle b, T_\rho b \rangle$ for all $b
\in {\cal B}$. Since sesquilinear forms are uniquely determined by
their quadratic forms we find $D_\rho(b_1, b_2) =
\langle b_1, T_\rho b_2 \rangle$ for all $b_1, b_2
\in {\cal B}$ for which the supports of $b_1$ and $b_2$ are equal.
Let, finally, $e
:= 1$ denote the indifferent proposition which is always true,
then $T_\rho$ satisfies $\langle e, T_\rho e \rangle = 1$.

This shows that the history version of standard quantum mechanics
over a finite dimensional single time Hilbert space ${\frak H}$
can indeed be brought into the form of a space time quantum theory as
formulated in Section III.A. Here the propositions Hilbert space
${\frak K}$ is independent of the initial quantum state $\rho$ and
only the ``temporal'' quantum state $T_\rho$ depends on $\rho$.

In the operator formulation of the history version of quantum mechanics
propositions are identified with projection operators and
\emph{consistent sets} of propositions are defined with the help of
$d_\rho$ as those exhaustive sets ${\cal C}$
of mutually perpendicular projections
for which Re $d_\rho(p,q) = 0$ for all $p,q \in {\cal C}$. Comparing
this with our definition of consistent sets of propositions in the
propositions Hilbert space ${\frak K}$ given in Section III.A
we see that -- in the case of finite dimensional
standard quantum mechanics --
every consistent set of propositions in
${\frak K}$ corresponds -- when pulled back to the standard operator
formulation -- to a consistent set of projection operators.

Given now some proposition $\bar{p} \in {\frak K}$ corresponding on the
operator level to some projection operator $p$. The larger the
space on which $p$ projects is,
the larger is also the norm of the image $\bar{p}$ of $p$ in
${\frak K}$. This
 substantiates our physical interpretation of $\Vert b \Vert_2^2 =
\langle b, b \rangle$ as a quantitative measure of how coarse grained
the proposition $b \in {\frak K}$ is.

We notice in passing
that for any $p \in \Bbb R$ with $p \geq 1$ we
can define a norm on $\cal B$ by \begin{equation}
\label{pnorm} \Vert b \Vert_p := \left(
\frac{{\rm tr}_{{\frak H}_{t_1} \otimes \cdots \otimes {\frak
H}_{t_n}} \left( (b^\dagger b)^{\frac{p}{2}}
\right)}{{\rm tr}_{{\frak H}_{t_1}
\otimes \cdots \otimes {\frak H}_{t_n}} \left( 1 \right)}
\right)^{\frac{1}{p}} \end{equation} for all $b \in \cal B$ where
$\{ t_1, \cdots, t_n \}$ denotes the support of $b$.
For a proof see for instance Schatten
\cite{Schatten70}, Section V.6. The norm $\Vert \cdot \Vert_2$
induced by the inner
product $\langle \cdot, \cdot \rangle$ obviously corresponds to
$p=2$. (Notice that $\Vert \cdot \Vert_p$ defines no crossnorm on
$\mathcal{B}$.)

Next consider
the case that the single time Hilbert space ${\frak H}$
is infinite dimensional. To simplify notation we assume in the sequel
that $\frak H$
is separable. The extension of our results to non separable
Hilbert spaces is obvious.
We proceed in analogy with the finite
dimensional case. The algebraic tensor product of
${\cal B}({\frak H}_{t_1}), \cdots,
{\cal B}({\frak H}_{t_n})$ is the set of all finite linear
combinations of homogeneous operators $b_1 \otimes \cdots \otimes b_n$,
where $b_i \in {\cal B}({\frak H}_{t_i})$ and is denoted by
${\cal B}({\frak H}_{t_1}) \otimes_{alg} \cdots \otimes_{alg}
{\cal B}({\frak H}_{t_n})$. Consider the directed system $\{
{\cal B}({\frak H}_{t_1}) \otimes_{alg} \cdots \otimes_{alg}
{\cal B}({\frak H}_{t_n}) \vert \{ t_1, \cdots, t_n \} \subset
{\Bbb R} \}$ and its direct limit which we denote
by ${\cal B}_{alg}$ (its existence as a $C^*$-algebra
follows again by Proposition
11.4.1 in \cite{KadisonR86}).
Define the map $\pi$ on homogeneous elements of ${\cal B}_{alg}$ as
in the finite dimensional case by $\pi(b_1 \otimes \cdots \otimes b_n) =
b_1 \cdots b_n$, then it follows by Proposition 11.1.8 in
\cite{KadisonR86} that $\pi$ can be uniquely extended to a linear map on
${\cal B}_{alg}$. Again we denote the extension of $\pi$
also by $\pi$ (slightly abusing the notation).
As in the finite dimensional case the decoherence functional $d_\rho$
associated with $\rho$ can be extended to a sesquilinear form defined
on all of ${\cal B}_{alg}$ and
the extension $D_\rho$ of $d_\rho$ can be written as
\[ D_\rho :
{\cal B}_{alg} \times {\cal B}_{alg} \to {\Bbb C}, D_\rho(b_1,
b_2) = \text{tr}_{{\frak H}}(\pi(b_1)^\dagger \rho
\pi(b_2)). \]
We remark that
the representation Equation (\ref{decf}) is also valid for $D_\rho$ on all
of ${\cal B}_{alg}$ (this follows by linearity).

Let $b_1, b_2 \in {\cal B}_{alg}$, then
we define \[ \langle b_1, b_2 \rangle =
\text{tr}_{{\frak H}_{t_1} \otimes \cdots \otimes {\frak H}_{t_n}}
(b_1^\dagger b_2), \] where $\{t_1, \cdots, t_n \}$ is the support of
$b_1^\dagger b_2$. This expression is not well defined
for arbitrary
$b_1$ and $b_2$. If it is not well
defined, then we formally set $\langle b_1, b_2 \rangle
:= \infty$.
It is clear that the elements $b \in {\cal B}_{alg}$ with finite norm
$\Vert b
\Vert^2 = \langle b, b \rangle < \infty$ are exactly the
Hilbert-Schmidt operators in ${\cal B}_{alg}$. In particular,
$\Vert b \Vert = 0$ for $b \in {\cal B}_{alg}$ implies $b =0$.
It is well known that every
trace class operator $\tau \in {\cal B}_{alg}$ satisfies
$\text{tr}((\tau^\dagger \tau)^{1/2}) < \infty.$ \\

We interpret $\langle \cdot, \cdot \rangle$ as an
improper inner product on ${\cal B}_{alg}$ (see Appendix A).
This implies in
particular that additivity is only required in the finite sectors
of ${\cal B}_{alg}$. Thus the factor $\text{tr}(1)$ appearing in
the definition of $\langle \cdot, \cdot \rangle$ in the finite
dimensional case is not only not well defined but also not needed
to ensure additivity in the finite sectors of ${\cal B}_{alg}$.
The
completion of ${\cal B}_{alg}$ with
respect to $\langle \cdot, \cdot \rangle$ is an improper Hilbert space
which we denote by ${\frak K}$. We interpret ${\frak K}$
as in the
finite dimensional case as our propositions Hilbert space.
[The main reason for completing ${\cal B}_{alg}$ here is to get a
mathematically nicer space of propositions.] We
find that the decoherence functional $d_\rho$ associated with $\rho$ can
be uniquely extended to a bounded sesquilinear form $\widehat{D}_\rho$
on ${\frak K}$. To see that ${D}_\rho$ is indeed bounded with
respect to the norm induced by the inner product $\langle \cdot, \cdot
\rangle$, recall the Cauchy-Schwarz inequality for $D_\rho$ \[
\vert D_\rho(b_1, b_2) \vert^2 \leq D_\rho(b_1, b_1) D_\rho(b_2, b_2), \]
for all $b_1, b_2 \in {\cal B}_{alg}$.
When $b \in {\cal B}_{alg}$
is a Hilbert-Schmidt operator with $n$-time support, it follows that
there is a constant $C>0$ such that
\[ \vert
D_\rho(b,b) \vert \leq C  \Vert b \Vert_{HS}^2, \]
where $\Vert \cdot \Vert_{HS}$ denotes the Hilbert-Schmidt norm.
To see this, recall the representation Equation (\ref{decf}) and
apply the Cauchy-Schwarz inequality
\begin{eqnarray}
\lefteqn{\vert D_\rho(b,b) \vert} & & \nonumber \\
& \leq &
\sum_{j_1, \cdots, j_{2n}=1}^{\infty} \omega_{j_1}
\left\vert
\left\langle e^{2n}_{j_{2n}} \otimes \cdots \otimes e^{n+1}_{j_{n+1}}
\otimes \psi_{j_1} \otimes \cdots \otimes e^n_{j_n},
(b^\dagger \otimes b)
(\psi_{j_1} \otimes e^{2n}_{j_{2n}} \otimes \cdots \otimes
e^{n+2}_{j_{n+2}} \otimes e^2_{j_2} \otimes \cdots \otimes
e^{n+1}_{j_{n+1}}) \right\rangle \right\vert, \nonumber \\
& \leq &
\sum_{j_1, \cdots, j_{2n}=1}^{\infty}
\left\vert
\left\langle e^{2n}_{j_{2n}} \otimes \cdots \otimes e^{n+1}_{j_{n+1}}
\otimes \psi_{j_1} \otimes \cdots \otimes e^n_{j_n},
(b^\dagger \otimes b)
(\psi_{j_1} \otimes e^{2n}_{j_{2n}} \otimes \cdots \otimes
e^{n+2}_{j_{n+2}} \otimes e^2_{j_2} \otimes \cdots \otimes
e^{n+1}_{j_{n+1}}) \right\rangle \right\vert, \nonumber \\
& \leq & \sum_{j_1, \cdots, j_{2n}=1}^{\infty}
\left\vert
\left\langle e^{2n}_{j_{2n}} \otimes \cdots \otimes e^{n+1}_{j_{n+1}}
\otimes \psi_{j_1} \otimes \cdots \otimes e^n_{j_n},
(b^\dagger b \otimes b b^\dagger)
(e^{2n}_{j_{2n}} \otimes \cdots \otimes
e^{n+1}_{j_{n+1}} \otimes \psi_{j_1} \otimes \cdots \otimes
e^{n}_{j_{n}}) \right\rangle \right\vert^{\frac{1}{2}}, \nonumber \\
& = & \sum_{j_1, \cdots, j_{n}=1}^{\infty}
\left\vert
\left\langle \psi_{j_1} \otimes \cdots \otimes e^{n}_{j_{n}},
(b^\dagger b) (\psi_{j_1} \otimes \cdots \otimes
e^{n}_{j_{n}}) \right\rangle \right\vert \nonumber \\
& = & \Vert b \Vert_{HS}^2 \nonumber \\
& = & \Vert [b]^2 \Vert_{1} \nonumber
\end{eqnarray}

\noindent
where $\Vert \cdot \Vert_{1}$ denotes the trace class norm and $[b] :=
(b^\dagger b)^\frac{1}{2}$.
>From the definition of a Cauchy sequence (see Appendix A)
it follows that for every Cauchy
sequence $\{ u_n \}$ there exists an $N$ such that $n, m > N$
implies that
$u_n - u_m$ is a Hilbert-Schmidt operator and $[u_n - u_m]^2 =
(u_n - u_m)^\dagger (u_n -
u_m)$ is a trace class operator converging to 0 in trace class norm. Thus
it follows from the above inequalities that $D_\rho$ can be uniquely
extended to a finitely valued sesquilinear form $\widehat{D}_\rho$
on ${\frak K}$.

We denote the subset of ${\frak K}$ of all elements with finite norm by
${\frak K}_{fin}$. The space ${\frak K}_{fin}$
is a union of proper Hilbert spaces (the Hilbert spaces of Hilbert-Schmidt
operators with fixed temporal support).
Consequently there exists a bounded operator $\widehat{T}_{\rho, i}$
on each Hilbert space ${\frak K}_i \subset {\frak K}_{fin}$
such that $\widehat{D}_\rho(x,x) = \langle x,
\widehat{T}_{\rho, i} x \rangle$, for all $x \in {\frak K}_{i}$.

The sesquilinear form $\widehat{D}_\rho$
also satisfies $\widehat{D}_\rho(e, e) = 1$, where $e
:= 1$ again denotes the indifferent proposition which is always true.

Summarizing we have shown that also the history version of standard
quantum mechanics over an infinite dimensional Hilbert space can be
brought into the form of a temporal quantum theory.

The reader might wonder whether the history version of standard
quantum mechanics can be brought into the form of a temporal
quantum theory with a proper propositions Hilbert space. The
answer is yes with the restriction that the sesquilinear forms
(which are the states in the present framework) are then either
only defined on a dense subset of the propositions Hilbert space
or coincide with the inner product of the propositions Hilbert
space. The latter natural representation of the standard
decoherence functional (in infinite dimensions) as an inner
product of a Hilbert space has been derived in
\cite{RudolphW98}. In this formulation all probabilistic
information is completely encoded within the inner product of
the propositions Hilbert space and there is no additional
notion of a state.
Accordingly, the information entropy to be defined in the next
section is always zero for this representation.
Otherwise, every positive linear
functional $\phi$ with $\phi(1) = 1$ induces
a semi inner product on ${\cal B}_{alg}$ by $\langle x, y
\rangle_\phi = \frac{1}{2} \phi(y^\dagger x + x y^\dagger)$.
It has been shown in \cite{RudolphW98} that $D_\rho$ is
not bounded with respect to the norm induced by $\langle \cdot, \cdot
\rangle_\phi$ (see also Appendix B) and that $D_\rho$ is
unbounded with respect to any $C^*$-norm on ${\cal B}_{alg}$.
Thus $D_\rho$ cannot be extended to the Hilbert
space completion of ${\cal B}_{alg}/N_\phi$
with respect to $\langle \cdot,
\cdot \rangle_\phi$, where $N_\phi$ denotes the null space of $\langle
\cdot, \cdot \rangle_\phi$. Moreover, in general
the set $N_\phi$ may contain
physical histories with non-vanishing probability. Therefore for
this construction to make sense one has to ensure that $N_\phi$
contains no elements with non-vanishing probability. For details
the reader is referred to \cite{RudolphW98}.

\section{Information-entropy}
In this section we study the problem of defining an information entropy
within our framework of temporal quantum theories. We adopt the point of
view that -- loosely speaking --
the information entropy measures the lack of information and
is a quantitative measure of the total amount of missing information on
the ultramicroscopic structure of the system. \\

The problem of defining an information entropy for temporal quantum
theories was first addressed in the framework of Isham's history quantum
theories by Isham and Linden \cite{IshamL97}.
They restricted themselves, however, to
history theories over finite dimensional Hilbert spaces. They
considered the case that the space of histories is given by the set
${\cal P}({\cal H})$ of projections on some finite dimensional
Hilbert space ${\cal H}$ and that the state is given by some bounded
decoherence functional on ${\cal P}({\cal H})$. Recall that to
every decoherence functional $d$ there is a unique trace class operator
${\frak X}_d$ on ${\cal H} \otimes {\cal H}$ such that Equation
(\ref{ILS2}) holds.
They proceeded as follows: first they observed that there seems to be no
straightforward simple way to generalize the expression for the
information entropy in single time quantum mechanics $I_{s-t} = -
\text{tr}_{{\frak H}} (\rho \ln \rho)$ to history quantum theories
since ${\frak X}_d$ is in general neither self-adjoint nor positive.
Thus they defined in a first step an information entropy with respect to
a consistent set of histories (a \emph{window}) $W$ by replacing the
decoherence functional $d$ by another decoherence functional $d_W$ such
that $d_W$ coincides with $d$ on $W$ and such that the operator
${\frak X}_{d_W}$ associated with $d_W$ is self-adjoint and positive.
The information entropy with respect to $d$ and $W$ was defined as \[
I_{d,W} := - \text{tr}({\frak X}_{d_W} \ln {\frak X}_{d_W}) -
\ln \dim {\cal H}^2. \]
The term $- \ln \dim {\cal H}^2$ is added to ensure that the
information entropy is invariant under refinement.
Isham and Linden also showed that $I_{d,W}$ decreases (or remains
constant) under consistent
fine graining of $W$.
An information entropy $I_d$ associated with $d$ can then be defined by
\[ I_d := \min_W I_{d,W}, \] where the minimum is taken over all
consistent sets $W$ of $d$. There are alternative possibilities to
define an information entropy, see \cite{IshamL97}.
One important feature of the information entropy $I_{d,W}$ with respect
to $d$ and the window $W$ such defined is that
its definition involves explicitly the dimension of the underlying
history Hilbert space ${\cal H}$ and the dimension
of the projections in
$W$. Thus the definitions of $I_{d,W}$ and $I_d$
have no straightforward finite
extensions to infinite dimensional history Hilbert
spaces.

It is the purpose of this section to define a corresponding notion of
information entropy for our scheme of space time quantum theories using
the techniques described by Isham and Linden.
Consider a space time quantum theory as in Section III.A over
some Hilbert space ${\frak K}$ of propositions and some state
given by the Wright operator $T \in
{\cal W}_e({\frak K})$. Since $T$ is not
positive, in general the expression $-
\text{tr}_{{\cal H}}(T \ln T)$ is not well defined .
We proceed in analogy with Isham and Linden and pick some
set $W = \{x_i\}_{i \in I}$ of propositions in ${\frak K}$ which is
consistent with respect to $T$.
We define a positive self-adjoint operator $\widetilde{T}_W$ by \[
\widetilde{T}_W := \sum_{i \in I} \frac{\langle x_i, T x_i
\rangle}{\langle x_i, x_i \rangle} P_i, \]
where $P_i$ denotes the projections in ${\frak K}$ onto the subspace
spanned by $x_i$. The operator $\widetilde{T}_W$ is again a state
operator in ${\cal W}_e({\frak K})$, i.e.,
satisfies $\langle e, T e \rangle = 1$. To see this, we recall that $e =
\sum_{i \in I} x_i$. Thus, $\langle e, \widetilde{T}_W
e \rangle = \sum_{m,l} \langle
x_m, \widetilde{T}_W x_l \rangle = \sum_m \langle x_m, \widetilde{T}_W
x_m \rangle = \sum_m \langle x_m, T x_m \rangle$, where we have used
that $\langle x_m, \widetilde{T}_W x_m \rangle = \langle x_m, T x_m
\rangle$ for $x_m \in W$. But since $\{ x_i \}_{i \in I}$ is a
consistent set for $T$, it follows that $\sum_m \langle x_m, T x_m
\rangle = \langle e, T e \rangle = 1$. Thus $\langle e, \widetilde{T}_W
e \rangle = 1$ and $\widetilde{T}_W \in {\cal W}_e({\frak K})$.
For $\widetilde{T}_W$ the expression $-
\text{tr}_{{\cal H}}(\widetilde{T}_W \ln \widetilde{T}_W)$ is well
defined and this motivates the definition of the information entropy for
the state $T$ and the window $W$ \begin{equation} I_{T,W} := -
\text{tr}_{{\cal H}}(\widetilde{T}_W \ln \widetilde{T}_W) = -
\sum_{i \in I} \langle x_i, T x_i \rangle \ln \frac{\langle x_i, T x_i
\rangle}{\langle x_i, x_i \rangle}. \label{ent} \end{equation}
An argument as in \cite{IshamL97} shows that $I_{T,W}$ decreases or
remains constant under refinements as it should. To this end,
we first notice that for $1 \leq q < \infty $\begin{equation}
a \ln \left(\frac{a}{b^q} \label{fab}
\right) - (1+a) \ln \left( \frac{(1+a)}{(1+b)^q} \right) \geq 0
\end{equation} for all $0 \leq a < \infty$ and $0< b < \infty$. To see
this let $f_q(a,b) \equiv
a \ln \left(\frac{a}{b^q}
\right) - (1+a) \ln \left( \frac{(1+a)}{(1+b)^q} \right)$.
The function $b \mapsto
f_q(a,b)$ assumes for every fixed $0 < a < \infty$ a minimum at $b =
a$. The value of this
minimum satisfies $f_q(a,a) \geq 0$ for all
$0 < a < \infty$ which proves the inequality (\ref{fab}). Now consider a
window $W_1 = \{ x_0, x_1, x_2, \cdots, x_n \}$ and a refinement $W_2 =
\{y_0, z_0, x_1, x_2, \cdots, x_n \}$ of $W_1$ where $x_0 = y_0 +
z_0$.
We define $a := \frac{\langle z_0, T z_0 \rangle}{\langle y_0, T y_0
\rangle}$ and $b := \frac{\langle z_0, z_0 \rangle}{\langle y_0, y_0
\rangle}$. A straightforward computation shows \[ I_{T, W_1} - I_{T,
W_2} = \langle y_0, T y_0 \rangle \left[ a \ln \left(\frac{a}{b}
\right) - (1+a) \ln \left( \frac{(1+a)}{(1+b)} \right) \right] \geq 0,
\] where we have used that that $\langle
x_0, x_0 \rangle = \langle y_0, y_0 \rangle + \langle z_0, z_0 \rangle$
since $W_2$ is a consistent set for $T$.
Thus the information entropy for $T$ and $W$ decreases (or remains
constant)
under any refinement of the window.

The information entropy for the Wright operator
$T$ can then be defined as the
minimum over all consistent sets, i.e., \[ I_T := \min_W I_{T, W}, \]
where the minimum is over all consistent sets of $T$.

It is instructive to compare the expression for the information entropy
$I_{T, W}$ for $T$ and $W$ given above
with the corresponding expression for the
Isham-Linden information entropy for a decoherence functional $d$ and a
window $V$ for $d$ in history quantum theories which was proposed in
\cite{IshamL97}
\[  I^{IL}_{d,V} := - \sum_{\alpha_i \in V} d(\alpha_i, \alpha_i) \ln
\frac{d(\alpha_i, \alpha_i)}{(\dim \alpha_i/ \dim {\cal H})^2}, \]
where ${\cal H}$ is the finite dimensional Hilbert
space on which the operators $\alpha_i$ act.
The factor $\dim {\cal H}$ is included to ensure the invariance of
the information entropy upon refinement of the consistent set.
Recalling that all $\alpha_i$ are projections, we see that the
expression for the Isham-Linden entropy can be written with the
norm $\Vert \cdot \Vert_{1}$ from Equation (\ref{pnorm})
as \[ I^{IL}_{d,V} := - \sum_{\alpha_i \in V} d(\alpha_i, \alpha_i) \ln
\frac{d(\alpha_i, \alpha_i)}{\Vert \alpha_i \Vert^2_{1}} =
- \sum_{\alpha_i \in V} \langle \alpha_i, T \alpha_i \rangle \ln
\frac{\langle \alpha_i, T \alpha_i \rangle}{\Vert \alpha_i \Vert^2_{1}}
\]
where $T$ is the Wright operator associated with $d$.
We see that there for any $1 \leq p < \infty$ there is
an Isham-Linden-type information entropy given by
\[ I^{IL}_{d,V,p} := - \sum_{\alpha_i \in V} d(\alpha_i, \alpha_i) \ln
\frac{d(\alpha_i, \alpha_i)}{\Vert \alpha_i \Vert^2_p}.
\]
All these expressions stand \emph{a priori} on an equal footing.
However, an argument as above shows that
$I^{IL}_{d,V,p}$ decreases or remains constant under
refinement of the consistent set if and only if $1 \leq p \leq 2$.
The proof is analogous to the
proof given above for the information entropy $I_{T,W}$ and makes
use of the general inequality (\ref{fab}).
Obviously the information entropy $I_{T,W}$ defined above in
Equation (\ref{ent})
corresponds to $p=2$
and the Isham-Linden entropy $I^{IL}_{d,V}$ corresponds to $p=1$.
The case $p=2$ is somewhat preferred since only in
this case the general construction given in Section III applies.

In the case of the history version of standard quantum mechanics over
infinite dimensional Hilbert spaces we see that the expression for the
information entropy $I_{T, W}$ might become infinite when the window $W$
involves coarse grained propositions $u$ with $\langle u, u \rangle =
\infty$. When we recall that the information entropy is a measure for the
amount of missing information, it is perhaps not too surprising that in
the infinite dimensional case (corresponding to an infinite variety of
possible measurement outcomes) the missing information becomes infinite
for certain windows involving `too' coarse grained propositions.
An alternative approach is (somewhat in the spirit of the topos
theoretic approach to the histories approach put forward by Isham
\cite{Isham96}) to define the information entropy by
\[ \widetilde{I}_{T, W} := \sup_{W_0} I_{T, W_0}, \] where
the supremum runs
over all consistent refinements
$W_0$ of $W$ such that $I_{T, W_0}$ is finite.
Notice, however, that $\widetilde{I}_{T, W}$ might also be
infinite.
\section{Summary}
In this paper we have put forward a mathematical framework for
temporal quantum theories involving observables associated with
extended regions of space time. The main ingredients of the
framework is a Hilbert space ${\frak K}$ which contains the
physical (measurable) propositions about the system. The norm of
an element in ${\frak K}$ is interpreted as a quantitative
measure of the structural information about the corresponding proposition
encoded within the space ${\frak K}$ and, more specifically,
as a quantitative
measure of the coarse grainedness of the corresponding
proposition within the descriptive scheme provided by
${\frak K}$. There is one distinguished element $e$ in ${\frak K}$
identified with the completely indifferent proposition which is
always true. The states are given by bounded,
self-adjoint, but
not necessarily positive
operators $T$ on ${\frak K}$ such that $\langle e, T e \rangle =
1$. The expression for the
probability of a proposition $x \in {\frak K}$ is given
by $\langle x, T x \rangle$ provided $x \in {\frak K}$.
This prescription makes sense when
one adopts a consistent-histories-type point of view according to
which the assignment of a probability to a proposition $x$
is unambiguously possible only
with respect to a consistent set of propositions containing $x$.

Our proposal is motivated by recent developments in the so-called
histories approach to quantum mechanics and we have seen that the
history version of standard quantum mechanics can be
brought into the required form in the finite dimensional case. In
the infinite dimensional case one has to allow for a slightly more
general framework in which the propositions Hilbert space ${\frak K}$
is an improper Hilbert space or -- alternatively -- in which the
states are given by densely defined unbounded sesquilinear forms
on the propositions Hilbert space.

We have also seen that Isham's general history quantum theories
can be brought into the form of a temporal quantum theory.
Moreover, we have defined an information entropy, generalizing the
Isham-Linden information entropy for history theories. \\

The examples discussed in Section III.B make clear that our
approach is not in contradiction to the history approach by Isham
et al.~but rather (in a sense) a complementary formulation of
temporal quantum theories.
In the case of standard quantum mechanics
we still can think of the space of propositions
${\frak K}$ essentially
as a set of operators on tensor product Hilbert spaces.
In this sense our approach may -- loosely speaking --
be looked upon as a compromise between
the formulations of history quantum theories due to Gell-Mann and Hartle
on the one hand and Isham on the other hand.

However, as already discussed above the history theory due to
Gell-Mann and Hartle stays essentially on the level of homogeneous
histories and does represent only a very modest generalization of
standard quantum mechanics. Isham's abstract history quantum
theories represent a much more substantial generalization of
standard quantum mechanics. However, it is an open problem, if and
how standard quantum mechanics can be recovered from them in some
appropriate limit. Specifically, it is not clear at all in which
limit a Hamiltonian operator can be recovered within the framework
of an abstract history theory. In contrast to these two
developments the approach developed in the present paper does offer
a generalization of standard quantum mechanics for which there is
hope that the issue of recovering standard quantum mechanics can
be successfully
tackled. A possibility suggesting itself is, for example, to
study propositions Hilbert spaces carrying a unitary representation
of the Poincar\'e group in which case a Hamiltonian operator can be
obtained as one of the generators of the representation. These
topics will be discussed elsewhere.

\begin{appendix}
\section{Improper Hilbert spaces}
Consider a vector space ${\frak V}$ equipped with an improper inner
product $\langle \cdot, \cdot \rangle_{{\frak V}} : {\frak V}
\times {\frak V} \to {\Bbb C} \cup \{ \infty \}$ such that (i)
$\langle w, a u + b v \rangle_{\frak V} = a \langle w, u
\rangle_{\frak V} + b \langle w, v \rangle_{\frak V}$, (ii)
$\langle u,v \rangle_{\frak V} = \langle v, u
\rangle_{\frak V}^*$, (iii) $\langle u,u \rangle_{\frak V} \geq 0$
and (iv) $\langle u,u \rangle_{\frak V} =0$ only if $u=0$, for all
$a, b \in {\Bbb C}$ and $u,v,w \in {\frak V}$ whenever all
expressions are finite. We denote the subspace of elements in ${\frak V}$
with finite norm by ${\frak V}_{fin}$.
A sequence $\{ u_n \vert n \in {\Bbb N},
u_n \in {\frak V} \}$ converges to $u \in {\frak V}$ if $\langle
u_n - u, u_n - u \rangle_{\frak V} \to 0$. A sequence
$\{ u_n \vert n \in {\Bbb N},
u_n \in {\frak V} \}$
is a \emph{Cauchy sequence} if  $\langle
u_n - u_m, u_n - u_m \rangle_{\frak V} \to 0$. Notice that for
any Cauchy sequence $\{ u_n \}$
there is an $N$ such that $n,m > N$ implies
$u_n - u_m \in {\frak V}_{fin}$. The space
${\frak V}$ is said to be
\emph{complete} if every Cauchy sequence converges.
An orthonormal basis is a set $\{ y_i \vert i \in {\Bbb I}, y_i \in
{\frak V} \}$ such that (i) $\langle y_i, y_j \rangle_{\frak V} =
\delta_{ij}$ for all $i,j$, (ii) $\langle u, y_i
\rangle_{\frak V} < \infty$ for all $i$ and $u \in {\frak V}$, and
(iii) $\langle u, y_i \rangle = 0$ for all $i$ if and only if $u=0$.
An \emph{improper Hilbert space} is now a linear space ${\frak V}$
with improper inner product $\langle \cdot, \cdot \rangle_{\frak V}$
such that (i) ${\frak V}$ is complete, and (ii) ${\frak V}$ has an
orthonormal basis $\{ y_i \}$.
Then every element $u \in  {\frak V}$ can be
formally expanded as $u = \sum_i
\langle u, y_i \rangle_{\frak V} y_i$. In contrast to ordinary
Hilbert spaces, however, the sum $\Vert u \Vert = \sum_i \vert \langle
u, y_i \rangle_{\frak V} \vert^2$ does not converge for all $u \in
{\frak V}$. We do not want to develop here a theory of improper
Hilbert spaces, but it is important to notice that many results of the
theory of Hilbert spaces are not valid for improper Hilbert spaces.
Notice, however, that there always resides some (non unique)
proper Hilbert space within an improper Hilbert space.
\section{The decoherence functional in standard quantum mechanics}
In \cite{RudolphW98} Wright and the author studied the analytical
properties of the standard decoherence functional $d_\rho$
associated with the initial state $\rho$. Among others we proved that if
the single time Hilbert space is infinite dimensional, then (i) the
standard decoherence functional $d_\rho$ defined on homogeneous
histories by Equation (\ref{decf1}) cannot be extended to a
finitely valued functional on the set of all projection operators
on the tensor product Hilbert space and (ii) the extension $D_\rho$ of
$d_\rho$ to ${\cal B}_{alg}$ is unbounded with respect to any $C^*$-norm
on ${\cal B}_{alg}$. The latter assertion together with Theorem
4.3.2 in \cite{KadisonR86} implies that $D_\rho$ is also unbounded
with respect to the norm induced by the inner product $\langle \cdot, \cdot
\rangle_\phi$ defined at the end of Section III [Theorem 4.3.2 in
\cite{KadisonR86} states that every positive linear functional $\phi$ on
${\cal B}_{alg}$ is bounded with respect to any $C^*$-norm on
${\cal B}_{alg}$]. We are not going to reproduce the general
considerations undertaken in \cite{RudolphW98} here but for the
convenience of the reader we give two counterexamples showing (i) and (ii)
respectively. We assume for simplicity that the single time
Hilbert space is separable. \\
(i) Consider the representation in Equation (\ref{decf}). For
simplicity of notation we consider the case $n=2$.
\begin{equation} \label{decf3} D_\rho(p,q) =
\sum_{j_1, \cdots, j_{4}=1}^{\dim {\frak H}}
\omega_{j_1}
\left\langle e^{4}_{j_{4}} \otimes e^3_{j_3}
\otimes \psi_{j_1} \otimes e^2_{j_2},
(p \otimes q) (\psi_{j_1} \otimes e^{4}_{j_{4}} \otimes
e^2_{j_2} \otimes e^3_{j_3})
\right\rangle,  \end{equation} for all
histories $p, q \in {\cal P}({\frak H}_{t_1} \otimes
{\frak H}_{t_2})$ for which the sum
converges. We assume that the single time Hilbert space ${\frak H}_t$
is separable. Now choose $e^4_j = e^3_j = e^2_j = \psi_j$ for all $j$. Fix
$i_1$ and let $\varphi_i := \frac{1}{\sqrt{2}} \left(
\vert \psi_i \otimes \psi_{i_1} \rangle + \vert \psi_{i_1} \otimes \psi_i
\rangle \right)$
for every $i \in {\Bbb N} \backslash \{ i_1 \}$.
Then clearly $\varphi_i \perp \varphi_j$ if $i \neq j$.
Set $f_{j_1,j_2,j_3}(q) = \langle \psi_{j_1} \otimes \psi_{j_2},
q (\psi_{j_2} \otimes \psi_{j_3}) \rangle$, then an easy
computation shows that \[ D_\rho(P_{\varphi_i}, q) = \frac{1}{2}
\sum_{j_2} \left( \omega_{i_1} f_{i_1,j_2,i_1}(q) + \omega_i
f_{i,j_2,i}(q) \right), \] for $i \neq i_1$ where $P_{\varphi_i}$
denotes the
projection operator onto the subspace spanned by $\varphi_i$. Put
$P=\sum_{i \neq i_1} P_{\varphi_i}$, then clearly the expression in
Equation (\ref{decf3}) for $D_\rho(P,q)$ does not
converge for arbitrary $q$. \\
(ii) Consider again $n=2$ and the operator \[ h = \sum_{k_1 k_4}
\frac{1}{k_1 + k_4} \vert e^4_{k_4} \otimes \psi_{k_1} \rangle
\langle \psi_{k_1} \otimes e^4_{k_4} \vert. \]
Then $h$ is a compact operator in the completion of the
algebraic tensor product ${\cal K}({\frak H}_{t_1}) \otimes_{alg}
{\cal K}({\frak H}_{t_2})$ [a Cauchy sequence $\{ h_n \}$ in
${\cal K}({\frak H}_{t_1}) \otimes_{alg}
{\cal K}({\frak H}_{t_2})$ converging to $h$ is for example
given by $h_n = \sum_{l=2}^n \sum_{k_1 k_4 \atop k_1 + k_4 = l}
\frac{1}{k_1 + k_4} \vert e^4_{k_4} \otimes \psi_{k_1} \rangle
\langle \psi_{k_1} \otimes e^4_{k_4} \vert$. Then $\Vert h_n - h_m \Vert
\leq \max ( \frac{1}{n}, \frac{1}{m})$].
Moreover, the sum in Equation
(\ref{decf3}) for $D_\rho(h,1)$ is equal to $\sum_{k_1, k_4}
\frac{\omega_{k_1}}{k_1 + k_4}$ and thus
clearly divergent. This shows that the canonical extension $D_\rho$
of $d_\rho$ on ${\cal B}_{alg}$ is not bounded on
${\cal K}({\frak H}_{t_1}) \otimes_{alg}
{\cal K}({\frak H}_{t_2})$ with respect to the ordinary
operator norm. Since, by nuclearity, all $C^*$-norms
on ${\cal K}({\frak H}_{t_1}) \otimes_{alg}
{\cal K}({\frak H}_{t_2})$ coincide, $D_\rho$
is unbounded with respect to any $C^*$-norm on
${\cal B}_{alg}$.

\end{appendix}

\subsection*{Acknowledgements}
The author is a Marie Curie Research Fellow and carries
out his research at Imperial College as part of a European
Union training project financed by the European Commission
under the Training and Mobility of Researchers (TMR)
programme. The author would like to thank
Professor Christopher J.~Isham for
reading a previous draft of this paper and
Professor John D.~Maitland Wright without whose
stimulating remarks and questions
the present work would not have been written.

\end{document}